\pdfoutput=1
\documentclass{article}

\usepackage{spconf, amsthm, amssymb}
\usepackage{bm}
\usepackage{algorithm}
\usepackage{color, latexsym}
\usepackage{amsmath, amsfonts}
\usepackage{subfigure}
\usepackage{tikz, pgfplots}
\usepackage[justification=centering]{caption}
\usepackage[square, comma,numbers, sort&compress]{natbib}
\usepackage{mathrsfs}
\usepackage{graphicx}
\usepackage{setspace}
\usepackage{booktabs}
\usepackage{epstopdf}

\usetikzlibrary{shapes.geometric, arrows}

\usepackage{algpseudocode} 

\def \x {\bm{x}}
\def \h {\bm{h}}
\def \y {\bm{y}}

\def \H {\bm{H}}
\def \X {\bm{X}}
\def \A {\bm{A}}
\def \p {\bm{p}}
\def \P {\bm{P}}
\def \G {\bm{G}}
\def \N {\bm{N}}
\def \Y {\bm{Y}}
\def \W {\bm{W}}
\def \Q {\bm{Q}}
\def \Z {\bm{Z}}
\def \R {\bm{R}}

\def \J {\bm{J}}

\newcommand{\myhat}[1]{\bm{\hat{#1}}}

\ninept

\theoremstyle{plain}

\theoremstyle{definition}

\theoremstyle{plain}

\theoremstyle{plain}

\theoremstyle{plain}

\DeclareMathOperator*{\minimize}{minimize}

\DeclareMathOperator*{\diag}{diag}

\title{Intelligent Reflecting Surface for Massive Device Connectivity: Joint Activity Detection and Channel Estimation}

\name{Shuhao Xia and Yuanming Shi}
\address{School of Information Science and Technology, ShanghaiTech University, Shanghai 201210, China\\
        Email: \{xiashh, shiym\}@shanghaitech.edu.cn
        }

\begin{document}
\topmargin=0mm

\maketitle

\begin{abstract}
    Intelligent Reflecting Surface (IRS) has been a promising solution to enhance wireless networks both spectral-efficiently and energy-efficiently. This paper considers an IRS-assisted the Internet of Things network for massive connectivity. We aim to solve the \emph{IRS-related activity detection and channel estimation} problem which has not been studied before. In this paper, we formulate the IRS-related activity detection and channel estimation problem as sparse matrix factorization, matrix completion and Multiple Measurement Vector problem and, we propose a three-stage framework based on the approximate message passing. Simulation results verify the effectiveness of the proposed algorithm.
\end{abstract}

\section{Introduction}
Intelligent Reflecting Surface (IRS) has recently emerged as a promising new technology for enhancing wireless communications both spectral-efficiently and energy-efficiently by controlling the propagation environment \cite{8796365, 8466374, zhangrui1}. Specifically, IRS is a planar surface consisting of a massive number of passive and programmable elements reflecting incident signals with phase shifts \cite{zhangrui1}. By smartly tuning the phase shifts, an IRS is able to reconfigure propagation environments for constructive signal combination and interference cancellation at the receivers, thereby enhancing the communication performance \cite{zhangrui1}.

To fully explore the benefits of IRS, the acquisition of the channel state information (CSI) plays a critical role in passive beamforming. \cite{8683663, yuanxiaojun2} proposed to estimate IRS-related channels based on training signals sent by transmitter or receiver. Assuming the perfect CSI, \cite{8811733,8683145} studied the problem of minimizing the base station (BS) transmit power by jointly optimizing the active beamforming at the BS and passive beamforming at the IRS. Furthermore, IRS has been jointly designed with other existing technologies, e.g., non-orthogonal multiple access (NOMA) \cite{fumin1} and over-the-air computation \cite{jiang2}. In particular, the IRS-related channel capacity characterization is studied in \cite{zhangrui3}.  However, there's no work on the use of IRS for providing massive device connectivity for the Internet of Things (IoT). In this paper, we propose to equip the IoT network with an IRS in order to support massive device connectivity. 

Massive device connectivity has been identified as one of the three main use cases in the upcoming 5G network, along with enhanced mobile broadband and ultra-reliable, low-latency communications \cite{6824752, 8808168}. In such a scenario, a large number of mobile devices are connected to the Internet via the BS with sporadic communications, e.g., only a small fraction of connected devices are active \cite{7047686}. To overcome the challenge of detecting active devices and estimating their channels, \cite{weiyu1,7852531} studied the activity detection and channel estimation problem for massive connectivity from the view of information theory. By exploiting the sparse activity pattern, the problem is formulated as a compressed sensing problem and resolved by the approximate message passing (AMP) \cite{8264818,8323218}. However, all these works treated the communication channels as an uncontrollable environment, and in some cases, harsh propagation environments can significantly degrade the system performance \cite{8466374}. Hence, we shall propose an IRS-assisted IoT network for massive connectivity to improve propagation environment.

In this paper, we consider an uplink IRS-assisted IoT network, where a single BS serves a massive number of mobile devices with the assist of an IRS. Specifically, our goal is to jointly detect active devices and estimate the IRS-related channels. We call this problem as the \emph{IRS-related activity detection and channel estimation} problem. In fact to our best knowledge, the problem has not been studied in the prior works. Due to unfavorable propagation conditions, the direct link between the BS and the devices has negligible received signals and thus we ignore the device-BS channel \cite{8741198, 8796365, 8461496}. However, deployment of an IRS poses new challenges, e.g., passive elements on the IRS can not process incident signals and there are more links to estimate \cite{8683663, yuanxiaojun2, zhangrui3}. To overcome these challenges, we formulate the IRS-related activity detection and channel estimation problem as sparse matrix factorization \cite{5197422, 8764368}, matrix completion and Multiple Measurement Vector (MMV) problem \cite{4014378}. To solve the problem, we propose a three-stage framework based on the approximate message passing (AMP) including the BiG-AMP algorithm \cite{6898015} for sparse matrix factorization, the Singular Value Thresholding (SVT) algorithm \cite{doi:10.1137/080738970} for matrix completion and the Vector AMP algorithm \cite{8264818} for the MMV problem. Simulation results demonstrate that the propose algorithm can achieve IRS-related device activity detection and channel estimation for an IRS-assisted ToT network.

\emph{Notations.}  $\mathbb{R}$ ($\mathbb{C}$) denotes the set of real (complex) numbers. $\X_{i,j}$ is the $(i,j)$-th entry of a matrix $\X$, and the operation $\diag\{\x\}$  with $\x\in\mathbb{C}^{n}$ returns a diagonal matrix $\X\in\mathbb{C}^{n\times n}$ where $\X_{i,i} = \x_i$. $(\cdot)^\top, \|\cdot\|_*, \odot, \mathcal{CN}(0,1)$ denote transpose operation, nuclear norm, the Hadamard product and standard complex Gaussian distribution. 

\begin{figure}[htbp]
    \setlength{\abovecaptionskip}{-0.cm}
    \setlength{\belowcaptionskip}{-0.cm}
    \centering
    \includegraphics[width=1\linewidth]{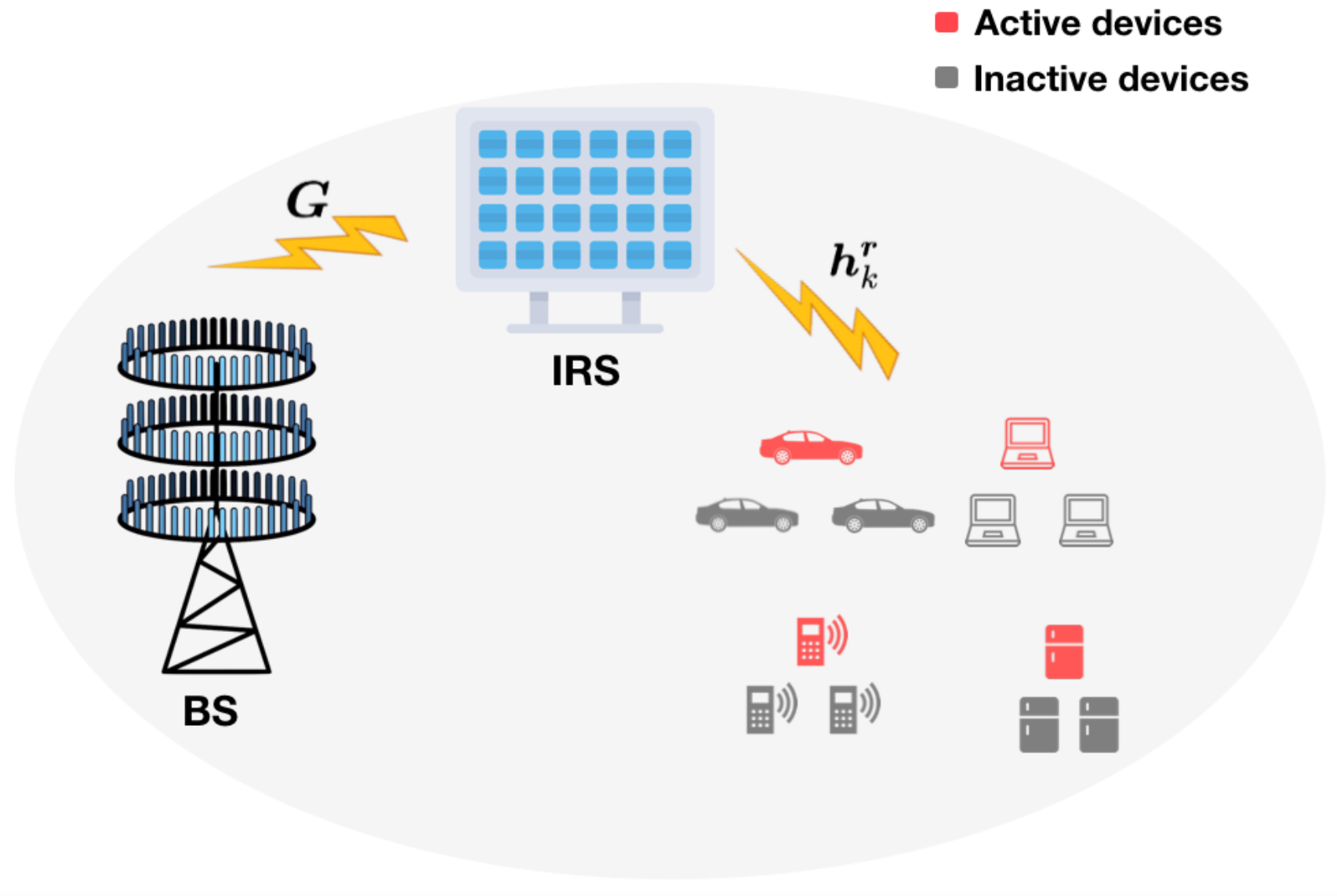}
    \caption{IRS-assisted IoT network without device-BS channels}
    \label{system model diagram}
\end{figure}

\section{System Model}
Consider an IRS-assisted IoT network in Fig \ref{system model diagram}, where the IRS consists of $N$ reflecting elements and the BS is equipped with $M$ antennas to serve $K$ single-antenna mobile devices. Due to unfavorable propagation conditions, the direct link between the BS and the devices has negligible received signals and thus we ignore the device-BS channel \cite{8741198, 8796365, 8461496}. We define $\bm{h}_k^r\in\mathbb{C}^{N\times 1}$ as the channel vector from the $k$-th device to the IRS. The channel matrix from the IRS to the BS is denoted by $\bm{G} \in \mathbb{C}^{M\times N}$. We assume a block-fading channel model where the channel is quasi-static in each block of length $T$. Furthermore, we consider Rayleigh fading and the path-loss fading for all the channels, e.g., $\bm{h}_k^r \sim \mathcal{CN}(\bm{0}, l_k\bm{I})$ and $\bm{G}\sim\mathcal{CN}(\bm{0}, l_{\text{R}}\bm{I})$, where $l_k$ and $l_{\text{R}}$ denote the path-loss components.

In this paper, we focus on sporadic communications \cite{7047686,8536396}, e.g., only a subset of all devices are active in each coherence block with probability $\rho$ and $\alpha_k = 1$ denotes the active state of the $k$-th device, otherwise $\alpha_k = 0$. For the propose of IRS-related device activity detection and channel estimation, the $k$-th device will transmit a unique signature sequence $\bm{x}_k = \left[\bm{x}_k(1), \cdots, \bm{x}_k(L)\right]\in \mathbb{C}^{1\times L}$ where $L < T$ is the sequence length. Then, for all $l=1,\cdots,L$,  the uplink signal received by the BS can be written as
\begin{align}\label{system model 1}
    \y(l) = \G\left(\p(l)\odot \sum_{k=1}^K\h_k^r\alpha_k \x_k(l)\right)
    +\bm{n}(l),
\end{align}
where $\bm{n}(l)\sim \mathcal{CN}(0, \sigma^2\bm{I})$ is the additive noise at time slot $l$, and the variance $\sigma^2$ depends on the background noise power normalized by the device transmit power. $\p(l) = \left[\beta_{l,1} e^{\jmath\phi_{l,1}},\cdots,\beta_{l,N} e^{\jmath\phi_{l,N}}\right]^\top \in \mathbb{C}^{N\times 1}$ is defined as the phase-shifting vector of the IRS, where $\beta_{l,j} \in \{0,1\}$ and $\phi_{l,j} \in (0,2\pi]$ are on/off state  and the phase shift of the $j$-th reflecting element on the IRS at the time slot $l$, respectively.

Let $\bm{H} = [\bm{h}^r_1,\cdots, \bm{h}^r_K] \in \mathbb{C}^{N\times K}$ and $\bm{X} = [\bm{x}_1^\top, \cdots, \bm{x}_K^\top]^\top \in \mathbb{C}^{K\times L}$ be the channel matrix of all device-IRS links and the known signature matrix, respectively. Considering $L$ time slots in each coherence block, the received signals $\Y \in \mathbb{C}^{M\times L}$ by the BS can be expressed in matrix form as 
\begin{align}\label{system model 2}
    \Y = \G\left(\P\odot\left(\H\A\X\right)\right) + \N,
\end{align}
where $\bm{N} = \left[\bm{n}(1)\cdots,\bm{n}(L)\right] \in \mathbb{C}^{M\times L}$ is the additive Gaussian noise, $\bm{A} = \diag\{\alpha_1,\cdots,\alpha_K\} \in \mathbb{R}^{K\times K}$ and  $\P = \left[\p_1(l), \cdots, \p_L(l)\right]\in \mathbb{C}^{N\times L}$ denote the activity matrix and the phase-shifting matrix, respectively.

This paper focuses on the regime where the number of devices is much larger than the signature sequence length, i.e., $K \gg L$. Therefore, it is impossible to assign the mutually orthogonal signature sequences to all devices. Inspired by \cite{8264818}, the signature sequence of the $k$-th device is generated from i.i.d complex Gaussian distribution with zero mean and variance one, i.e., $\x_k \sim \mathcal{CN}(0,\bm{I}_L)$.

\section{Problem Formulation}
The goal of this paper is to jointly detect the activity matrix $\A$ and estimate the IRS-related channel matrices $\G$ from the IRS to the BS and $\H$ from the active devices to the IRS, given the received signals $\Y$, the phase-shifting matrix $\P$ and the known signature matrix $\X$ in the regime where $K\gg L$. We call this problem as the \emph{IRS-related activity detection and channel estimation} problem. For simplicity, we introduce the following variables
\begin{align}
    \bm{\Theta} &= \bm{HA}\in\mathbb{C}^{N\times K}, \label{active channels} \\
    \Q &= \bm{\Theta}\X \in \mathbb{C}^{N\times L}, \label{Q} \\
    \W &= \P\odot\Q \in \mathbb{C}^{N \times L}. \label{W}
\end{align}
Since the activity matrix $\A$ is a sparse diagonal matrix, $\bm{\Theta}$ has the group sparsity on its column \cite{matrin1}. Then, the system model (\ref{system model 2}) can be expressed as
\begin{align} \label{sparse matrix factorization}
    \Y = \G\W + \N.
\end{align}
From the definition (\ref{W}), we can see that the matrix $\W$ has the same sparsity pattern as the phase-shifting matrix $\P$.  Hence, by designing the matrix $\P$ as a sparse matrix, we can  recover the matrix $\G$ and the matrix $\W$  from the observations $\Y$ via the techniques of \emph{sparse matrix factorization} \cite{5197422, 8764368}. Specifically, we design the phase-shifting matrix $\P$ as follows: the on/off state $\beta_{l,j}$ of the $j$-th reflect element on the IRS at the time slot $l$ are generated independently from Bernoulli distribution with the probability $\lambda$ of taking the value $1$. In addition, we generate the phase shifts $\phi_{l,j}$ according to the standard uniform distribution within $(0, 2\pi]$.

Due to the sparsity of the matrix $\W$ and (\ref{W}), we have to recover the missing entries of $\Q$ given the estimated matrix $\myhat{\W}$ and the phase-shifting matrix $\P$. Note that $\Q$ is low-rank due to the group sparsity of $\bm{\Theta}$. Hence, this sub-problem can be formulated as a \emph{matrix completion} problem and solved by exploiting the low-rank property of $\Q$.

Finally, recovering the $\bm{\Theta}$ from the estimated matrix $\myhat{\Q}$ and the known signature matrix $\X$ turns out to be the \emph{Multiple Measurement Vectors (MMV)} problem in compressed sensing \cite{4014378}.

In summary, the original problem can be solved through the following three stages:
\begin{enumerate}
    \item \textbf{Sparse Matrix Factorization:} Recovering the matrices $\G$ and $\W$ from the observations $\Y$;
    \item \textbf{Matrix Completion:} Completing the missing entries of $\Q$ given the estimated matrix $\myhat{\W}$ and the phase-shifting matrix $\P$;
    \item \textbf{Multiple Measurement Vectors:} Estimating the matrix $\bm{\Theta}$ from the estimated matrix $\myhat{\Q}$ and the known signature matrix $\X$.
\end{enumerate}

Once obtaining the estimated matrix $\myhat{\bm{\Theta}}$, the activity matrix $\myhat{\A} = \diag\{\hat{\alpha}_1, \cdots, \hat{\alpha}_K\}$ can be recovered via the group sparsity of $\bm{\Theta}$ as follows
\begin{align} \label{threshold}
    \myhat{\alpha}_k = \left\{\begin{array}{l}
            1, \|\myhat{\bm{\theta}}_k\|_2 >  \epsilon\\
            0,  \|\myhat{\bm{\theta}}_k\|_2 \leq  \epsilon
    \end{array}\right.\quad 1\leq k \leq K,
\end{align}
where $\epsilon$ is some small non-negative threshold and $\myhat{\bm{\theta}}_k$ is the $k$-th column of the estimated matrix $\myhat{\bm{\Theta}}$. Thus, the estimated matrix $\myhat{\H}$ can be estimated by setting its $i$-th column as $\myhat{\h}_i = \myhat{\bm{\theta}}_i$ where $i \in \{k|\hat{\alpha}_k = 1\}$ \cite{8536396}.

\section{Proposed IRS-Related Activity Detection and Channel Estimation Algorithm}
To solve IRS-related activity detection and channel estimation problem, we establish a three-stage framework based on the approximate message passing (AMP) algorithm. As shown in Algorithm \ref{alg:Framwork}, the architecture of overall algorithm consists of the BiG-AMP algorithm \cite{6898015} for sparse matrix factorization, the Singular Value Thresholding (SVT) algorithm \cite{doi:10.1137/080738970} for matrix completion and the Vector AMP algorithm \cite{8264818} for the MMV problem. We will explain the details of the proposed algorithm in the following subsections.
\begin{algorithm}[htbp]
    \setlength{\abovecaptionskip}{0.cm}
    \setlength{\belowcaptionskip}{-0.cm}
    \caption{Proposed Joint IRS-Related Activity Detection and Channel Estimation Algorithm}
    \label{alg:Framwork} 
    \begin{itemize}
    	\item \textbf{Stage 1. Sparse Matrix Factorization via BiG-AMP} (\ref{Stage 1})
    	\begin{algorithmic}[1]  
    		\Require  
    		The observations $\Y$.
    		\Ensure  
    		The estimated matrices $\myhat{\G}$ and $\myhat{\W}$.
    	\end{algorithmic} 
    	\item \textbf{Stage 2. Matrix Completion via SVT} (\ref{Stage 2})
    \begin{algorithmic}[1]  
    	\Require  
        The estimated matrix $\myhat{\W}$ in \textbf{Stage 1} and the phase-shifting matrix $\P$.
    	\Ensure  
    	The estimated matrix $\myhat{\Q}$. 
    \end{algorithmic} 
    	\item \textbf{Stage 3. Multiple Measurement Vector problem via Vector AMP} (\ref{Stage 3})
	\begin{algorithmic}[1]
		\Require  
        The estimated matrix $\myhat{\Q}$ in \textbf{Stage 2} and the known signature matrix $\X$.
		\Ensure  
		The estimated matrix $\myhat{\Theta}$.
	\end{algorithmic}  
    \end{itemize}
 \end{algorithm}

 \subsection{Sparse Matrix Factorization via BiG-AMP} \label{Stage 1}

First, we adopt an extension of the AMP algorithm, the BiG-AMP algorithm \cite{6898015} to solve the following generalized bilinear inference problem: estimate matrices $\G=\left[g_{m,n}\right] \in \mathbb{C}^{M\times N}$ and $\W=\left[w_{n,l}\right]\in \mathbb{C}^{N\times L}$ from the observations $\Y=\left[y_{ml}\right] \in \mathbb{C}^{M\times L}$ according to (\ref{sparse matrix factorization}). The BiG-AMP algorithm solves the sparse matrix factorization problem by modeling $\G$ and $\W$ as random matrices. Specifically, we introduce the following \emph{maximum a posteriori} (MAP) estimation problem:
\begin{align}
    \left(\myhat{\G}, \myhat{\W}\right) &= \arg \max_{\G, \W} p\left(\G, \W | \Y\right) \nonumber \\
    &= \arg \max_{\G, \W} p\left(\Y| \Z\right) p\left(\G\right) p\left(\W\right), 
\end{align}
where $\Z = \G\W$ and we assume that the likelihood function is known and separable, i.e.,
\begin{align}
    p(\Y |\Z)&=\prod_{m} \prod_{l} p\left(y_{m l} | z_{m l}\right).
\end{align}
Furthermore, we assume the entries of $\G$ obey the i.i.d Gaussian prior and the entries of $\W$ obey the i.i.d zero-mean Bernoulli-Gasussian prior \cite{6898015}. Hence, the prior of $\G$ and $\W$ can be modeled as follows
\begin{align}
    p(\G) &= \prod_m\prod_n\mathcal{CN}(g_{m,n};0, \sigma_g), \\
    p(\W) &= \prod_n\prod_l(1-\lambda)\delta(w_{nl}) + \lambda\mathcal{CN}(w_{nl};0,\sigma_w), 
\end{align}
where $\lambda$ denotes the sparsity level of $\W$; $\sigma_g$ and $\sigma_w$ are the variances of $\G$ and $\W$, respectively. To achieve the estimation, the BiG-AMP algorithm infers $\G$ and $\W$ from the above model so that the corresponding estimated matrices $\myhat{\G}$ and $\myhat{\W}$ can be multiplied to yield an estimate $\myhat{\Y} = \myhat{\G}\myhat{\W}$ up to the permutation and phase ambiguities. Details of the BiG-AMP can be found in \cite{6898015}.


\subsection{Matrix Completion via SVT} \label{Stage 2}
After sparse matrix factorization and ambiguity elimination in section \ref{Stage 1}, we obtain the estimates $\myhat{\G}$ and $\myhat{\W}$. Due to the design of the phase-shifting matrix $\P$, we have to recover the missing entries in $\Q=\left[q_{nl}\right]\in\mathbb{C}^{N\times L}$ from the estimated matrix $\myhat{\W}$. According to (\ref{W}), $\W$ has the same sparsity pattern as $\P$. By exploiting the low-rank property of $\Q$, we shall solve the following optimization problem:
\begin{align}
    \minimize_{\Q\in\mathbb{C}^{N\times L}}\quad &\left\|\Q\right\|_*\\
    \text{subject to}\quad &\mathcal{P}_\Omega(\P \odot\Q) = \mathcal{P}_\Omega(\myhat{\W})
\end{align}
where $\Omega = \left\{(i,j)|\P_{ij} \neq 0 \right\}$. $\mathcal{P}_{\Omega}$ denotes an orthogonal projector onto the subspace spanned by matrices with sampled entries in $\Omega$, namely, the $(i,j)$-th entry in $\mathcal{P}_\Omega(\W)$ is equal to $\W_{i,j}$ if $(i,j) \in \Omega$ and zero otherwise. To solve such a matrix completion problem, we apply the Singular Value Thresholding (SVT) algorithm \cite{doi:10.1137/080738970}. Specifically, for a constant $\tau$ and a sequence $\left\{\delta_k\right\}_{k\geq 1}$, starting with $\J^0 = 0$, the iterations of SVT proceed as
\begin{align}
\left\{\begin{array}{l}{\Q^{k}=\mathcal{S}_{\tau}\left(\J^{k-1}\right)} \\ {\J^{k}=\J^{k-1}+\delta_{k} \mathcal{P}_{\Omega}\left(\myhat{\W}-\P\odot\Q^{k}\right)}\end{array}\right.,
\end{align}
where $\mathcal{S}_{\tau}\left(\J^{k-1}\right)$ is a soft-thresholding operator at level $\tau$ to the singular values of the input matrix, which is defined as  $\mathcal{S}_{\tau}(\J) =\boldsymbol{U} \operatorname{diag}\left(\left\{(\sigma_{i}-\tau)_{+}\right\}\right) \boldsymbol{V}^{*}$ and $(\cdot)_+$ is the positive part of the input, e.g., $(t)_+ = \max(0, t)$. 

\begin{figure*}[htbp]
    \setlength{\abovecaptionskip}{0.cm}
    \setlength{\belowcaptionskip}{-0.cm}
	\centering
	\subfigure[]{
		\begin{minipage}[t]{0.33\linewidth}\label{nmse of G}
			\centering
            \includegraphics[width=1.1\linewidth]{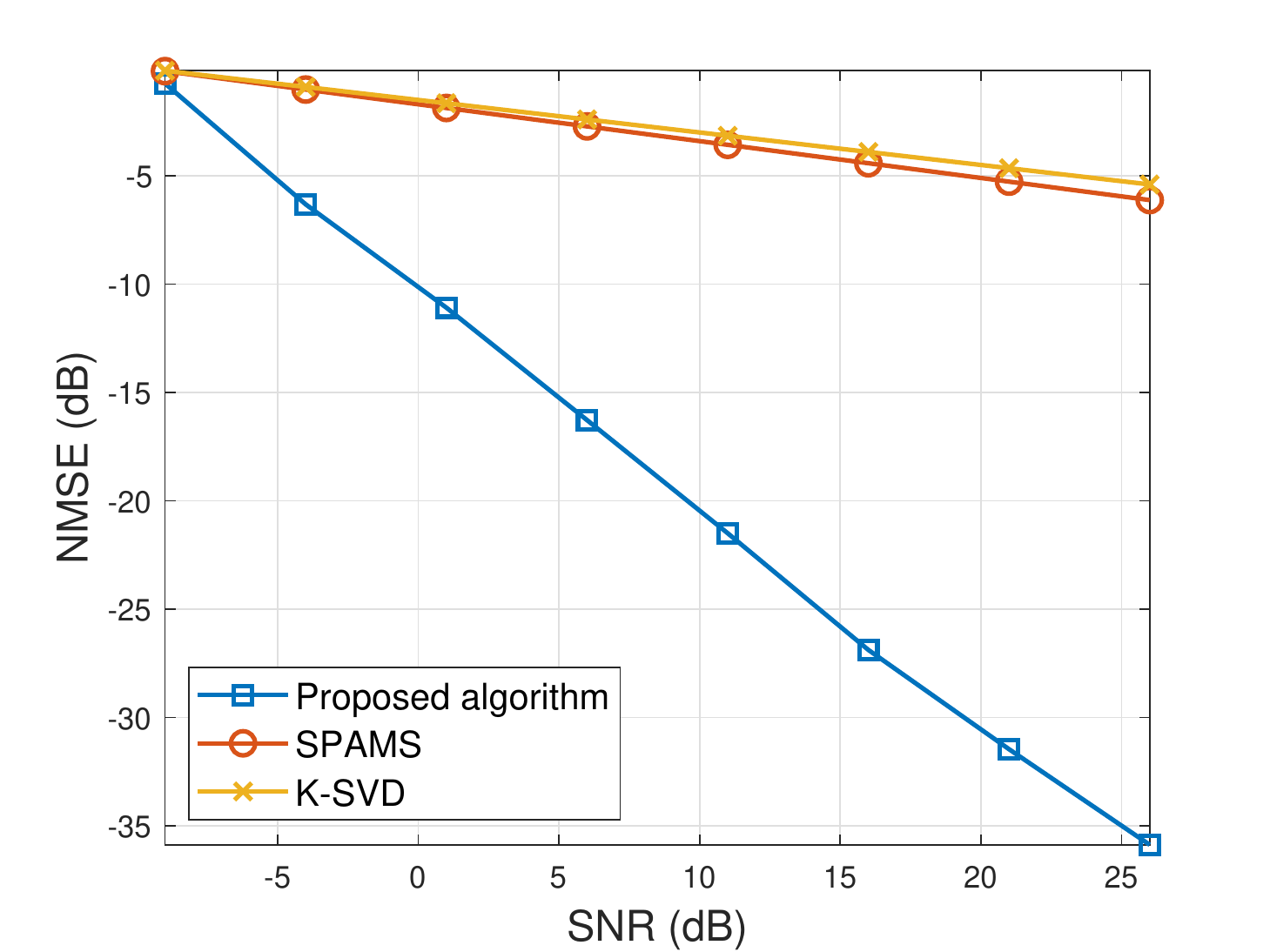}
		\end{minipage}%
	}%
	\subfigure[]{
		\begin{minipage}[t]{0.33\linewidth}\label{error rate}
			\centering
            \includegraphics[width=1.1\linewidth]{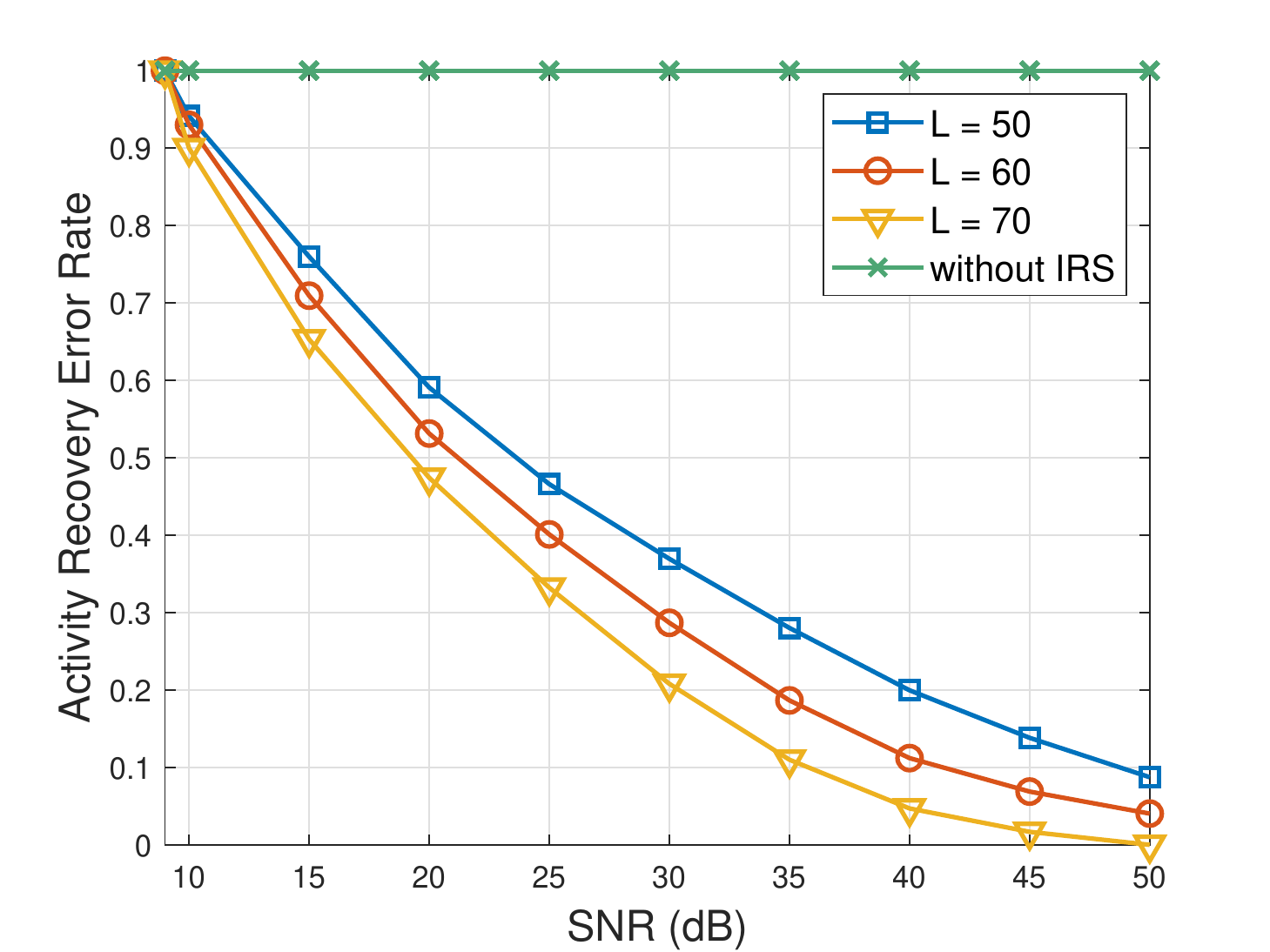}
		\end{minipage}
    }%
    \subfigure[]{
		\begin{minipage}[t]{0.33\linewidth}\label{phase transition}
			\centering
            \includegraphics[width=1.1\linewidth]{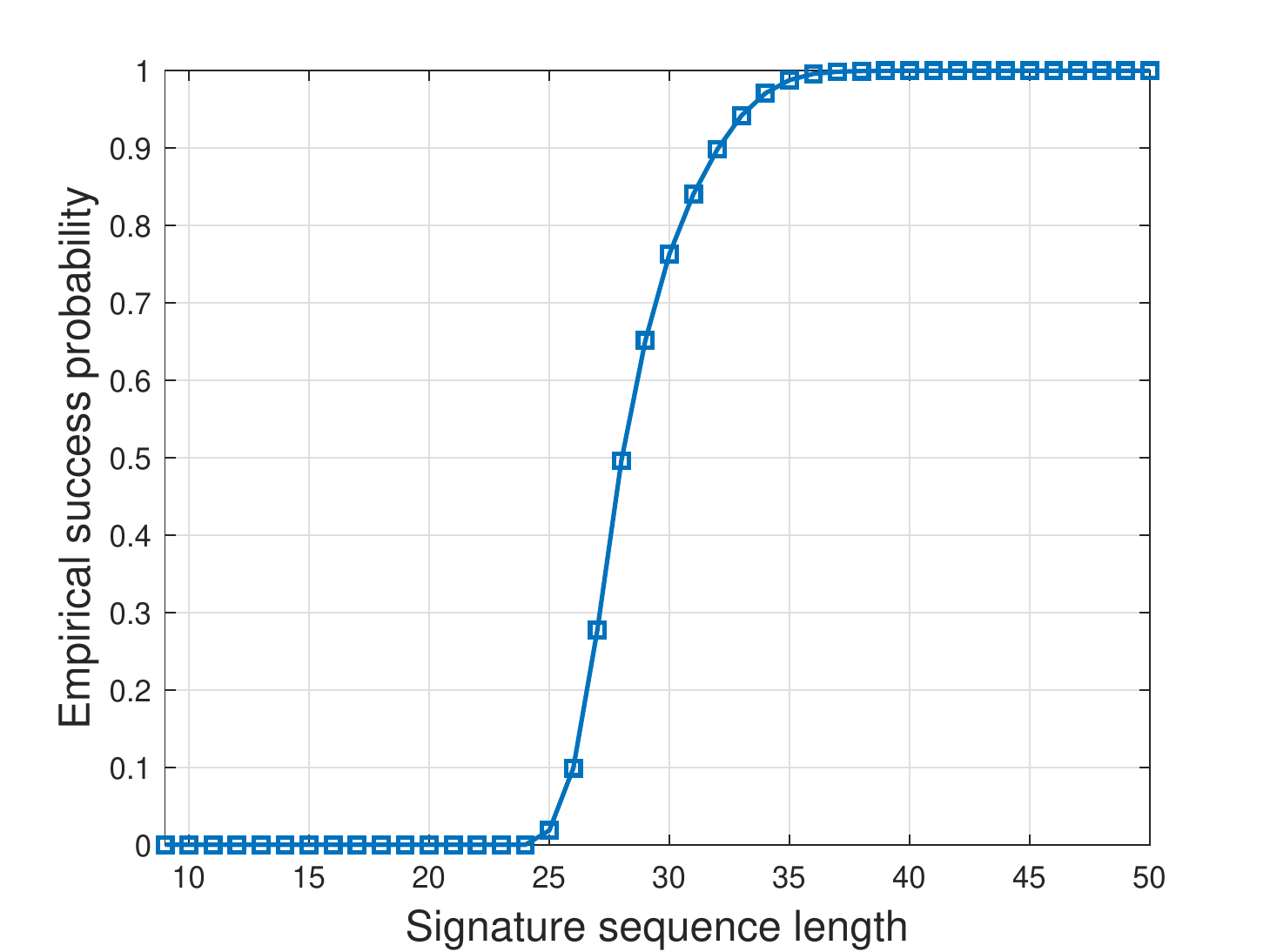}
		\end{minipage}
	}%
    \caption{Simulation results}
\end{figure*}

\subsection{MMV problem via Vector AMP} \label{Stage 3}
As for the MMV problem of recovering $\myhat{\bm{\Theta}}$ from the estimated matrix $\myhat{\Q}$ and the known signature matrix $\X$, we adopt the Vector AMP algorithm proposed in \cite{Belief}, which operates a vector denoiser on each column vector of the matching filter output. initializing with ${\myhat{\Theta}}^0 = 0$ and $\R^0={\Q}$, the iterations of the vector AMP algorithm are defined as
\begin{equation}\label{vec-amp}
\begin{aligned}
    \myhat{\theta}^t_k &= \left((\R^t)^*{\X} + {\myhat{\Theta}}^{t}\right)_k, \\
{\myhat{\theta}}^{t+1}_k &= \eta_{t,k}(\myhat{\theta}^t_k), \\
\R^{t+1} &= {\Q} - {\X}{\myhat{\Theta}}^{t+1} + \frac{K}{L}\R^t\sum_{k=1}^K\frac{\eta'_{t,k}(\myhat{\theta}^t_k)}{K},
\end{aligned}
\end{equation}
where $\myhat{\theta}^t_k$ is the $k$-th column vector of the estimated matrix ${\myhat{\Theta}}^t$ at iteration t, $\R^t \in \mathbb{C}^{N\times L}$ is the corresponding residual, $\eta_{t,k}(\cdot): \mathbb{C}^{N\times 1}\rightarrow\mathbb{C}^{N\times 1}$ is a vector denoiser that operates on the $k$-th column vector of $(\R^t)^*{\X} + {\myhat{\Theta}}^{t}$, and $\eta'_{t,k}(\cdot)$ is the first-order derivative of $\eta_{t,k}(\cdot)$. Here, we apply the MMSE denoiser derived in \cite{8323218}, and this denoiser has the following form:
\begin{align*}
    \eta_{t,k}(\myhat{\theta}^t) = \phi_{t,k} \frac{l_k}{l_k + \tau^2_t}\myhat{\theta}^t,
\end{align*}
where 
\begin{align*}
    \phi_{t, k}&=\frac{1}{1+\frac{1-\rho}{\rho} \exp \left(-\frac{N}{2}\left(\pi_{t, k}-\psi_{t, k}\right)\right)},\\
    \pi_{t, k}&=\left(\frac{1}{\tau_{t}^{2}}-\frac{1}{\tau_{t}^{2}+l_{k}}\right) \frac{(\myhat{\theta}^t)^{H} \myhat{\theta}^t}{N},\\
    \psi_{t, k}&=\log \left(1+\frac{l_{k}}{\tau_{t}^{2}}\right), \\
    \tau_{t+1} &= \sigma^2 + \frac{K}{L}\rho\mathbb{E}_l\left[\frac{l\tau_t^2}{l + \tau^2_t}\right],
\end{align*}
with $\tau_0 = \sigma^2 + \frac{K}{L}\rho\mathbb{E}_l[l]$ and $l$ denoting path-loss component.

\section{Simulation}
We carry out numerical experiments to evaluate the performance of the proposed algorithm for massive device connectivity with employment of IRS, where the direct path is ignored due to unfavorable propagation conditions \cite{8741198, 8796365, 8461496}.

For all the considered channels, we assume Rayleigh fading and the path loss \cite{8323218,8811733}. The path loss model is given by $l(d) = l_0(\frac{d}{d_0})^{-\alpha}$ where $l_0$ is the path loss at the reference distance $d_0$. In our setting, $d_0 = 1$, $l_0 = -30$dB, and the path loss component $\alpha$ for device-to-IRS link and IRS-to-BS link are set as 2 and 2.8, respectively \cite{8811733, jiang2}. The $k$-th device-to-IRS distance $d_k$ is randomly generated from 500m to 1000m and the IRS-to-BS distance $d_R$ is 100m. We suppose the noise power normalized by the device transmit power, and thus the signal-to-noise ratio (SNR) is defined as $10\log_{10}(\frac{1}{\sigma})$ where $\sigma^2$ is the variance of the noise.  We resolved the permutation and phase ambiguities based on the true values of $\G$ and $\W$. For all the simulations: the signature matrix $\X$ is generated from $\mathcal{CN}(0,\bm{I})$, and we set $N = 15$ IRS elements, $M = 30$ antennas at the BS and $K = 200$ devices among which each device is active with probability $\rho=0.05$, and the sparsity $\lambda$ of the phase-shifting matrix $\P$ is fixed at 0.5. 

The performance of recovering the IRS-related channels is evaluated in terms of normalized mean-square-error (NMSE). To benchmark the estimation of the IRS-related channel $\G$, we compare the proposed algorithm with other two algorithms of sparse matrix factorization, K-SVD \cite{1710377} and SPAMS \cite{Mairal}. In Fig.\ref{nmse of G}, we show that the proposed algorithm significantly outperforms other two baseline algorithms, which demonstrates the effectiveness of our algorithm.

We illustrate the activity recovery error rate versus the SNR for different values of the signature sequence length $L$. We can see that the larger $L$, the faster the error rate can be driven to zero as the SNR increases. As shown in Fig. \ref{error rate}, we can not recover any active device without the IRS under unfavorable propagation conditions, however, this issue can be resolved by deploying an IRS in IoT networks. Furthermore, we consider the noiseless case and study the empirical success probability of recovering the IRS-related channel $\H$ versus $L$. We declare successful recovery if the NMSE of $\H<-50$dB and record the success probability from 50 trials. The sharp phase transition result in Fig. \ref{phase transition} are thus able to guide the selection of the signature sequence length.

\section{Conclusion}
In this work, we considered the device activity detection and channel estimation problem for the IRS-assisted IoT network. We establish a three-stage framework including the BiG-AMP algorithm for sparse matrix factorization, the Singular Value Thresholding algorithm for matrix completion and the Vector AMP algorithm for the multiple measurement vector problem. To the end, we provided the simulation results to verify the effectiveness of the proposed algorithm.
\vfill\pagebreak
\footnotesize
\bibliographystyle{IEEEtran}
\bibliography{reference}

\begin{thebibliography}{10}
\providecommand{\url}[1]{#1}
\csname url@samestyle\endcsname
\providecommand{\newblock}{\relax}
\providecommand{\bibinfo}[2]{#2}
\providecommand{\BIBentrySTDinterwordspacing}{\spaceskip=0pt\relax}
\providecommand{\BIBentryALTinterwordstretchfactor}{4}
\providecommand{\BIBentryALTinterwordspacing}{\spaceskip=\fontdimen2\font plus
\BIBentryALTinterwordstretchfactor\fontdimen3\font minus
  \fontdimen4\font\relax}
\providecommand{\BIBforeignlanguage}[2]{{%
\expandafter\ifx\csname l@#1\endcsname\relax
\typeout{** WARNING: IEEEtran.bst: No hyphenation pattern has been}%
\typeout{** loaded for the language `#1'. Using the pattern for}%
\typeout{** the default language instead.}%
\else
\language=\csname l@#1\endcsname
\fi
#2}}
\providecommand{\BIBdecl}{\relax}
\BIBdecl

\bibitem{8796365}
E.~{Basar}, M.~{Di Renzo}, J.~{De Rosny}, M.~{Debbah}, M.~{Alouini}, and
  R.~{Zhang}, ``Wireless communications through reconfigurable intelligent
  surfaces,'' \emph{IEEE Access}, vol.~7, pp. 116\,753--116\,773, 2019.

\bibitem{8466374}
C.~{Liaskos}, S.~{Nie}, A.~{Tsioliaridou}, A.~{Pitsillides}, S.~{Ioannidis},
  and I.~{Akyildiz}, ``A new wireless communication paradigm through
  software-controlled metasurfaces,'' \emph{IEEE Commun. Mag.}, vol.~56, no.~9,
  pp. 162--169, Sep. 2018.

\bibitem{zhangrui1}
\BIBentryALTinterwordspacing
Q.~{Wu} and R.~{Zhang}, ``Towards smart and reconfigurable environment:
  Intelligent reflecting surface aided wireless network,'' \emph{IEEE Commun.
  Mag.}, Early Access. [Online]. Available:
  \url{https://arxiv.org/abs/1905.00152}
\BIBentrySTDinterwordspacing

\bibitem{8683663}
D.~{Mishra} and H.~{Johansson}, ``Channel estimation and low-complexity
  beamforming design for passive intelligent surface assisted miso wireless
  energy transfer,'' in \emph{Proc. IEEE Int. Conf. Acoust., Speech Signal
  Process. (ICASSP)}, Brighton, U.K., May 2019, pp. 4659--4663.

\bibitem{yuanxiaojun2}
\BIBentryALTinterwordspacing
Z.-Q. He and X.~Yuan, ``Cascaded channel estimation for large intelligent
  metasurface assisted massive mimo,'' \emph{CoRR}, vol. abs/1905.07948, 2019.
  [Online]. Available: \url{https://arxiv.org/abs/1905.07948}
\BIBentrySTDinterwordspacing

\bibitem{8811733}
Q.~{Wu} and R.~{Zhang}, ``Intelligent reflecting surface enhanced wireless
  network via joint active and passive beamforming,'' \emph{IEEE Trans.
  Wireless Commun.}, pp. 1--1, 2019.

\bibitem{8683145}
------, ``Beamforming optimization for intelligent reflecting surface with
  discrete phase shifts,'' in \emph{Proc. IEEE Int. Conf. Acoust., Speech
  Signal Process. (ICASSP)}, Brighton, U.K., May 2019, pp. 7830--7833.

\bibitem{fumin1}
M.~{Fu}, Y.~{Zhou}, and Y.~{Shi}, ``Intelligent reflecting surface for downlink
  non-orthogonal multiple access networks,'' in \emph{Proc. IEEE Global Commun.
  Conf. (Globecom) Workshops}, Hawaii, USA, Dec 2019.

\bibitem{jiang2}
\BIBentryALTinterwordspacing
T.~{Jiang} and Y.~{Shi}, ``Over-the-air computation via intelligent reflecting
  surfaces,'' in \emph{Proc. IEEE Global Commun. Conf. (Globecom)}, Hawaii,
  USA, Dec 2019. [Online]. Available: \url{https://arxiv.org/abs/1904.12475}
\BIBentrySTDinterwordspacing

\bibitem{zhangrui3}
\BIBentryALTinterwordspacing
S.~{Zhang} and R.~{Zhang}, ``Capacity characterization for intelligent
  reflecting surface aided mimo communication,'' \emph{CoRR}, vol.
  abs/1910.01573, 2019. [Online]. Available:
  \url{https://arxiv.org/abs/1910.01573}
\BIBentrySTDinterwordspacing

\bibitem{6824752}
J.~G. {Andrews}, S.~{Buzzi}, W.~{Choi}, S.~V. {Hanly}, A.~{Lozano}, A.~C.~K.
  {Soong}, and J.~C. {Zhang}, ``What will 5g be?'' \emph{IEEE J. Sel. Areas
  Commun.}, vol.~32, no.~6, pp. 1065--1082, June 2014.

\bibitem{8808168}
K.~B. {Letaief}, W.~{Chen}, Y.~{Shi}, J.~{Zhang}, and Y.~A. {Zhang}, ``The
  roadmap to 6g: Ai empowered wireless networks,'' \emph{IEEE Commun. Mag.},
  vol.~57, no.~8, pp. 84--90, August 2019.

\bibitem{7047686}
G.~{Wunder}, H.~{Boche}, T.~{Strohmer}, and P.~{Jung}, ``Sparse signal
  processing concepts for efficient 5g system design,'' \emph{IEEE Access},
  vol.~3, pp. 195--208, 2015.

\bibitem{weiyu1}
W.~Yu, ``On the fundamental limits of massive connectivity,'' in \emph{Proc.
  Inf. Theory Appl. Workshop}, San Diego, Ca, USA, Feb 2017, pp. 1--6.

\bibitem{7852531}
X.~{Chen}, T.~{Chen}, and D.~{Guo}, ``Capacity of gaussian many-access
  channels,'' \emph{IEEE Trans. Inform. Theory}, vol.~63, no.~6, pp.
  3516--3539, June 2017.

\bibitem{8264818}
Z.~{Chen}, F.~{Sohrabi}, and W.~{Yu}, ``Sparse activity detection for massive
  connectivity,'' \emph{IEEE Trans. Signal Process.}, vol.~66, no.~7, pp.
  1890--1904, April 2018.

\bibitem{8323218}
L.~{Liu} and W.~{Yu}, ``Massive connectivity with massive mimo—part i: Device
  activity detection and channel estimation,'' \emph{IEEE Trans. Signal
  Process.}, vol.~66, no.~11, pp. 2933--2946, June 2018.

\bibitem{8741198}
C.~{Huang}, A.~{Zappone}, G.~C. {Alexandropoulos}, M.~{Debbah}, and C.~{Yuen},
  ``Reconfigurable intelligent surfaces for energy efficiency in wireless
  communication,'' \emph{IEEE Trans. Wireless Commun.}, vol.~18, no.~8, pp.
  4157--4170, Aug 2019.

\bibitem{8461496}
C.~{Huang}, A.~{Zappone}, M.~{Debbah}, and C.~{Yuen}, ``Achievable rate
  maximization by passive intelligent mirrors,'' in \emph{Proc. IEEE Int. Conf.
  Acoust., Speech Signal Process. (ICASSP)}, Alberta, Canada, April 2018, pp.
  3714--3718.

\bibitem{5197422}
Y.~{Koren}, R.~{Bell}, and C.~{Volinsky}, ``Matrix factorization techniques for
  recommender systems,'' \emph{Computer}, vol.~42, no.~8, pp. 30--37, Aug 2009.

\bibitem{8764368}
H.~{Liu}, X.~{Yuan}, and Y.~J. {Zhang}, ``Super-resolution blind
  channel-and-signal estimation for massive mimo with one-dimensional antenna
  array,'' \emph{IEEE Trans. Signal Process.}, vol.~67, no.~17, pp. 4433--4448,
  Sep. 2019.

\bibitem{4014378}
J.~{Chen} and X.~{Huo}, ``Theoretical results on sparse representations of
  multiple-measurement vectors,'' \emph{IEEE Trans. Signal Process.}, vol.~54,
  no.~12, pp. 4634--4643, Dec 2006.

\bibitem{6898015}
J.~T. {Parker}, P.~{Schniter}, and V.~{Cevher}, ``Bilinear generalized
  approximate message passing—part i: Derivation,'' \emph{IEEE Trans. Signal
  Process.}, vol.~62, no.~22, pp. 5839--5853, Nov 2014.

\bibitem{doi:10.1137/080738970}
J.~Cai, E.~Candès, and Z.~Shen, ``A singular value thresholding algorithm for
  matrix completion,'' \emph{SIAM J. Optim.}, vol.~20, no.~4, pp. 1956--1982,
  2010.

\bibitem{8536396}
T.~{Jiang}, Y.~{Shi}, J.~{Zhang}, and K.~B. {Letaief}, ``Joint activity
  detection and channel estimation for iot networks: Phase transition and
  computation-estimation tradeoff,'' \emph{IEEE Internet Things J.}, vol.~6,
  no.~4, pp. 6212--6225, Aug 2019.

\bibitem{matrin1}
M.~Wainwright, ``Structured regularizers for high-dimensional problems:
  Statistical and computational issues,'' \emph{Annual Review of Statistics and
  Its Application}, vol.~1, pp. 233--253, 01 2014.

\bibitem{Belief}
\BIBentryALTinterwordspacing
J.~Kim, W.~Chang, B.~Jung, D.~Baron, and J.~C. Ye, ``Belief propagation for
  joint sparse recovery,'' \emph{CoRR}, vol. abs/1102.3289, 2011. [Online].
  Available: \url{https://arxiv.org/abs/1102.3289}
\BIBentrySTDinterwordspacing

\bibitem{1710377}
M.~{Aharon}, M.~{Elad}, and A.~{Bruckstein}, ``K-svd: An algorithm for
  designing overcomplete dictionaries for sparse representation,'' \emph{IEEE
  Trans. Signal Process.}, vol.~54, no.~11, pp. 4311--4322, Nov 2006.

\bibitem{Mairal}
J.~Mairal, F.~Bach, J.~Ponce, and G.~Sapiro, ``Online learning for matrix
  factorization and sparse coding,'' \emph{J. Mach. Learn. Res.}, vol.~11, pp.
  19--60, Mar. 2010.

\end{thebibliography}
\end{document}